\documentclass[pss]{wiley2sp}
\usepackage{amsmath}
\begin{document}
\title{Microwave-induced off-resonance giant magnetoresistance in ultraclean two-dimensional electron systems}

\titlerunning{Microwave-induced off-resonance giant magnetoresistance...  }

\author{Jesus I\~narrea}

\authorrunning{Jesus I\~narrea.}

\mail{e-mail
  \textsf{jinarrea@fis.uc3m.es}, Phone
  +34 91 334 624 9478 }

\institute{%
 Escuela Polit\'ecnica Superior,Universidad Carlos
III, 28911, Leganes, Madrid, Spain.}

\received{XXXX, revised XXXX, accepted XXXX} \published{XXXX}



\abstract{ \abstcol{%
We report on theoretical studies of a recently discovered strong
radiation-induced magnetoresistance spike obtained in ultraclean two-dimensional
electron systems at low
temperatures. The most striking feature of this strong spike is that it shows up
on the  second harmonic of the cyclotron resonance.}
{We apply  the radiation-driven electron orbits model
in the ultraclean scenario. Accordingly, we calculate the
new average advanced distance by the electron in a scattering event
which will define the unexpected resonance spike position.
Calculated results are in good agreement with experiments.}}
\maketitle
\section{ Introduction}
In the last decade it was discovered that when a high mobility 2DES in a low and perpendicular magnetic field ($B$)
is irradiated, mainly with microwaves (MW), some striking effects are revealed:
radiation-induced  magnetoresistance ($R_{xx}$) oscillations  and zero
resistance states (ZRS) \cite{mani1,zudov1}. Different
theories and experiments have been proposed to explain these  effects
\cite{ina2,ina20,girvin,dmitriev,lei,ryzhii,rivera,manifoton,shi} but the physical
origin is still being questioned.
One of the most challenging experimental
results, recently obtained\cite{yanhua,hatke,yanhua2,hatke2} consists in a strong resistance spike which shows up far off-resonance,
 at twice the
cyclotron frequency, $w\approx2w_{c}$\cite{yanhua,hatke},
were $w$ is the radiation
frequency and $w_{c}$ the cyclotron frequency.
Remarkably, the only different feature in these
experiments\cite{yanhua,hatke} is the use of ultraclean samples
with mobility $\mu\sim 3\times 10^{7}cm^{2}/V s$ and lower temperatures, $T\sim 0.4 K$.
Yet, for the previous, $"standard"$, experiments and samples\cite{mani1} the
mobility
is lower, $\mu < 10^{7}cm^{2}/V s$ and  $T$ higher, $T\geq 1.0 K$.
In this letter, we
theoretically study this radiation-induced
 $R_{xx}$ spike applying
the theory developed by the authors, {\it the radiation-driven electron
orbits model}\cite{ina2,ina20,ina3}. 
We show that, in the ultraclean regime,  the averaged measured current
results to be the same as the one obtained in
a sample with always full contribution to $R_{xx}$ but
as if it were irradiated with MW of   half frequency ($w/2$).
Accordingly, the
cyclotron resonance is apparently shifted to a new
$B$-position around $w\approx2w_{c}$.

\section{ Theoretical Model and Results}
\begin{figure}\includegraphics*[width=\linewidth,height=0.5\linewidth]{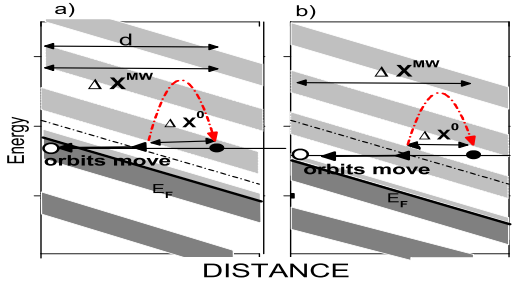}
\caption{Schematic diagrams presenting the two average electronic scattering scenarios for ultraclean samples.
In a),  the scattering ends around the central part of a LL (grey stripes).
In b) the
scattering jump ends in between LL (white stripes). The stripes represent
the $tilted$ Landau levels due to static electric field in the $x$ direction. $d$
is the inter-LL distance.
}
\end{figure}
The {\it radiation-driven electron orbits model}, was developed to explain
the $R_{xx}$ response of an irradiated 2DEG at low magnetic field\cite{ina2,ina20,ina3}.
The corresponding time-dependent
Schr\"{o}dinger equation can be exactly solved. We first obtain
an exact expression of the electronic wave vector:
$\Psi_{N}(x,t)\propto\phi_{n}(x-X-x_{cl}(t),t)$,
where $\phi_{n}$ is the solution for the
Schr\"{o}dinger equation of the unforced quantum harmonic
oscillator and 
$x_{cl}=\frac{e E_{o}}{m^{*}\sqrt{(w_{c}^{2}-w^{2})^{2}+\gamma^{4}}}\cos wt=A\cos wt$
where $E_{0}$ is the MW electric field. $\gamma$ is a  damping factor
for the electronic interaction with the lattice ions giving rise to emission of acoustic phonons.
\begin{figure}
\includegraphics*[width=\linewidth,height=0.6\linewidth]{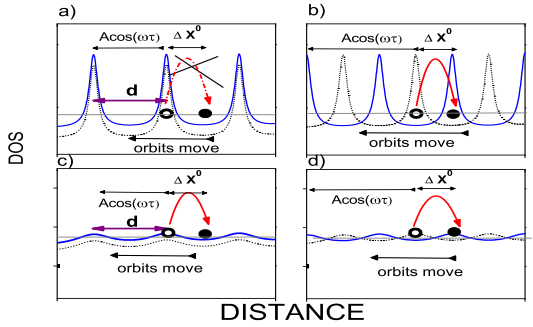}
\caption{Schematic diagrams of density of Landau states versus
energy for ultraclean, a) and b), and standard samples, c) and d). Dotted lines correspond to
the density of Landau states before the scattering and single lines
 after. We present for both kind of samples the 
 two scattering scenarios (see text).  $\Delta X^{0}$ is the advanced distance
in the dark.}
\end{figure}
\begin{figure}
\includegraphics*[width=\linewidth,height=0.4\linewidth]{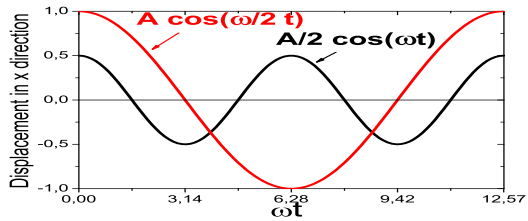}
\caption{Displacement of two harmonic motions vs time. One with half amplitude $A/2$ and frequency $w$ 
and the other with total amplitude $A$ and half frequency $w/2$. We represent two cycles 
observing that the net displacement covered by both is the same:$\left|\frac{A}{2}\cos w t\right|=\left|A\cos \frac{w}{2}t\right| $.
 }
\end{figure}
Thus, the electron orbit centers are not
fixed, but they oscillate harmonically at $w$.
 This $radiation-driven$ behavior will affect dramatically the
charged impurity scattering and eventually the conductivity.
We
 apply time dependent first order perturbation theory. This allows to
calculate the elastic charged impurity scattering rate
between two $oscillating$ Landau states, the initial $\Psi_{n}$ and the
final state $\Psi_{m}$ \cite{ina2,ina3}: $W_{n,m}=1/\tau$
being $\tau$ the elastic charged impurity scattering time.
Next, we find the average effective distance advanced by the electron
in every  scattering jump\cite{ina2,ina3},
$\Delta X^{MW}= \Delta X^{0}+A\cos w\tau$
where $\Delta X^{0}$ is the advanced distance
in the dark.  Finally the longitudinal conductivity
$\sigma_{xx}$ is given by
$\sigma_{xx}\propto \int dE \frac{\Delta X^{MW}}{\tau}$
being $E$
the energy.
After some algebra we obtain
that the dependence of the magnetoresistance
with radiation is mainly given by:
$R_{xx}\propto\frac{e
E_{o}}{m^{*}\sqrt{(w_{c}^{2}-w^{2})^{2}+\gamma^{4}}}\cos
w\tau$, 
\begin{figure}
\includegraphics*[width=\linewidth,height=0.7\linewidth]{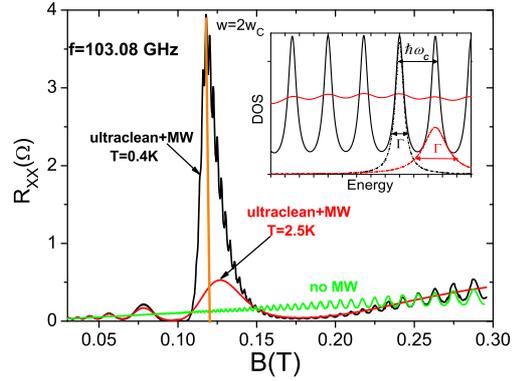}
\caption{Calculated irradiated  magnetoresistance vs static magnetic field for a radiation frequency of
$f=103.08 GHz$, for  a
ultraclean sample and temperatures, $T=0.4 K$ and $T=2.5 K$. We observe a intense spike at $w\approx2 w_{c}$.
In the inset, density of states for broad and narrow LL.}
\end{figure}

In standard samples (lower mobility) the LL width, $\Gamma$, is
large and
for experimental magnetic fields\cite{mani1}, it turns out that
$\Gamma> \hbar w_{c}$. In this regime due to the strong overlapping
of  LL, electrons always have an important density of available states where to get to after a
scattering event, either
around the center of LL or in between them.
For ultraclean samples $\Gamma$ is small and
generally,
$\Gamma< \hbar w_{c}$; the LL
hardly overlap each other leaving regions
with a low density of states in between (white stripes)
This condition will dramatically affect the
average advanced distance by electron in every scattering process.
Here,  we can clearly find
two opposite scenarios  described in Fig. 1.
In this figure, the
 grey stripes are LL tilted by the action of the DC electric field
in the x direction.  We
can observe regularly alternating grey (many states) and white (few states)
stripes equally spread out.
The first scenario corresponds  (see Fig. 1a) to an electron
being scattered  to the central part of a LL
where the scattering can be completed; we obtain an important contribution to
the conductivity and $R_{xx}$. In Fig. 1b, we describe the
second scenario where the electron scatters to a region
in between LL with a very low density of states. Obviously,
in this case there is no much contribution to the average 
stationary current.
In Fig. 2, we present schematic diagrams contrasting the different scattering scenarios for both
ultraclean and standard samples. Here, we present the
density of Landau states vs distance where dotted lines correspond to
the DOS before the scattering and single lines
the same but after the scattering. For all presented panels, during the scattering
time,  Landau states are driven backwards by the action of radiation. In 2a and 2b,
(ultraclean sample), the scattered electron can either
ends in between  LL  or around the center of one LL respectively.
In the first case we hardly obtain current; this happens when
$\Delta X^{MW} _{ultra}\simeq(n+\frac{1}{2})d$, where $d$ is the inter-LL distance (see Fig.1a and 2a).
In the second case we have full contribution and then the net 
advanced distance for ultraclean fulfills, $\Delta X^{MW}_{ultra}\simeq nd$.
In  average,
the two situations are equally distributed along one cycle of the
back and forth    MW-driven harmonic orbit motion.
In Figs. 2c and 2d we present the same  situation for
a standard sample. 
Now, for the two scattering scenarios
 we always obtain significant contribution to
$R_{xx}$. This condition can be mathematically expressed as   $\Delta X^{MW}_{stand}\simeq(n+\frac{1}{2})d+nd\simeq2nd$,
when $n>1$.
Comparing both, ultra (contributing part) and standard, we can write  $\Delta X^{MW}_{ultra}=\Delta X^{MW}_{stand}/2\propto\frac{A}{2}\cos w \tau$.
Then, regarding MW, the average advanced distance in ultraclean is equivalent to standard but
with half amplitude.
This point turns out to be crucial, because a 
harmonic motion with half amplitude $\frac{A}{2}$ and frequency  $w$
is in average physically equivalent to another one of amplitude $A$ and half
frequency $w/2$; first, the energy involved in both harmonic motions is the same and, what 
is more important, averaging in cycles the net displacement performed by both is
the same too (see Fig. 3). Thus, we can write:
 $\left|\frac{A}{2}\cos w t\right|=\left|A\cos \frac{w}{2}t\right| $.
Then, adapting this condition to our specific case of MW-driven  harmonic
motion:
$\left|\frac{A}{2}\cos w\tau\right|\simeq\left|A_{2}\cos \frac{w}{2}\tau\right| $
where $A\simeq A_{2}=\frac{e
E_{o}}{m^{*}\sqrt{(w_{c}^{2}-(\frac{w}{2})^{2})^{2}+\gamma^{4}}}$,
which is a good approximation according to experimental parameters\cite{yanhua}.
The consequence is that the
{\it ultraclean} harmonic motion behaves as if electrons
were driven by radiation  of half
frequency but with full amplitude (full to contribution to $R_{xx}$) during
the whole cycle.
Therefore, applying
next the theory\cite{ina2} for the ultraclean case it
is straightforward to reach  an expression
for magnetoresistance:
$R_{xx}\propto\frac{e
E_{o}}{m^{*}\sqrt{(w_{c}^{2}-(\frac{w}{2})^{2})^{2}+\gamma^{4}}}\cos
\frac{w}{2}\tau$
According to it, now the resonance in $R_{xx}$ will take place at
$w\approx2 w_{c}$, as experimentally obtained\cite{yanhua}.
The intensity of the $R_{xx}$ spike will depend on the relative value of
the frequency term, ($w_{c}^{2}-(\frac{w}{2})^{2}$), and the damping parameter $\gamma$ in
the denominator of the latter $R_{xx}$ expression.
When $\gamma$ leads the denominator the spike is smeared out. Yet,
in situations where $\gamma$ is smaller than the frequency term, the
resonance effect will be more visible and the spike will show up.
As we explained above, the parameter $\gamma$ represents the scattering
probability  of electrons with
the lattice ions and the release of
the radiation energy in form of acoustic phonons. According to Ando\cite{ando} $\gamma$ is
linear with the lattice temperature $T$:  $\gamma=\frac{1}{\tau_{ac}}\propto T$.
Now by decreasing $T$, it is possible
to reach a situation where $(w_{c}^{2}-(\frac{w}{2})^{2})^{2}>\gamma^{4}$ making visible a resonance effect and, therefore,  giving rise
to a strong resonance peak at $w\approx2 w_{c}$.
For GaAs and standard experimental parameters\cite{mani1}, we obtain that $\gamma\simeq 7-10 \times
10^{11} s^{-1}$. In a ultraclean experiments\cite{yanhua} it decreases till $\sim 3.5\times 10^{11}$.
This fixes a lower cut-off value for the radiation frequency where the resistance spike
could be observed. According to our calculations this would be
around $f=w/2\pi \approx 40-45 GHz$. 
In Fig. 4,  we present calculated irradiated $R_{xx}$ vs static magnetic field for
a radiation frequency of  $f=103.08 GHz$ and a ultraclean sample. The presented results correspond to  cases of
 high and low
temperatures. For $T=0.4 K$, we obtain a strong spike at $w\approx2 w_{c}$ as in
experiments\cite{yanhua}. Increasing temperature, ($T\simeq 2.5K$)  the spike vanishes but
the radiation-induced oscillations still remain but with less intensity.
\section{ Conclusions}
In this letter we have presented the first theoretical approach to the striking
result of the magnetoresistance spike in the second harmonic of the
cyclotron frequency. According to our model the strong change in the density 
of Landau states in ultraclean samples affects dramatically the electron impurity scattering 
and eventually the conductivity. The final results is that the scattered electrons perceive
radiation as of half frequency.
Calculated results are in good agreement with experiments.

This work is supported by the MCYT (Spain) under grant
MAT2011-24331 and ITN Grant 234970 (EU).

\end{document}